\documentclass[twocolumn]{aastex631}

\newcommand{\lir}{L$_{\mathrm{IR}}$}

\newcommand{\loglir}{$\log\mathrm{L}_{\mathrm{IR}}/\mathrm{L}_\odot$}

\newcommand{\lsun}{$\mathrm{L_\odot}$}

\begin{document}

\title{A Near-Infrared Faint, Far-Infrared-Luminous Dusty Galaxy at $z\sim5$ in COSMOS-Web}
\author[0000-0002-6149-8178]{Jed McKinney}
\affiliation{Department of Astronomy, The University of Texas at Austin, Austin, TX, USA}

\author[0000-0003-0415-0121]{Sinclaire M. Manning}
\altaffiliation{NASA Hubble Fellow}
\affiliation{Department of Astronomy, University of Massachusetts Amherst, 710 N Pleasant Street, Amherst, MA 01003, USA}

\author[0000-0003-3881-1397]{Olivia R. Cooper}\altaffiliation{NSF Graduate Research Fellow}
\affiliation{Department of Astronomy, The University of Texas at Austin, Austin, TX, USA}

\author[0000-0002-7530-8857]{Arianna S. Long}
\altaffiliation{NASA Hubble Fellow}
\affiliation{Department of Astronomy, The University of Texas at Austin, Austin, TX, USA}

\author[0000-0003-3596-8794]{Hollis Akins}
\affiliation{Department of Astronomy, The University of Texas at Austin, Austin, TX, USA}

\author[0000-0002-0930-6466]{Caitlin M. Casey}
\affiliation{Department of Astronomy, The University of Texas at Austin, Austin, TX, USA}

\author[0000-0002-9382-9832]{Andreas L. Faisst}
\affiliation{Caltech/IPAC, MS 314-6, 1200 E. California Blvd. Pasadena, CA 91125, USA}

\author[0000-0002-3560-8599]{Maximilien Franco}
\affiliation{Department of Astronomy, The University of Texas at Austin, Austin, TX, USA}

\author[0000-0003-4073-3236]{Christopher C. Hayward}
\affiliation{Center for Computational Astrophysics, Flatiron Institute, 162 Fifth Avenue, New York, NY 10010, USA}

\author[0000-0003-3216-7190]{Erini Lambrides}\altaffiliation{NPP Fellow}
\affiliation{NASA-Goddard Space Flight Center, Code 662, Greenbelt, MD, 20771, USA}

\author[0000-0002-4872-2294]{Georgios Magdis}
\affiliation{Cosmic Dawn Center (DAWN), Copenhagen, Denmark}
\affiliation{DTU-Space, Technical University of Denmark, Elektrovej 327, DK-2800 Kgs. Lyngby, Denmark}
\affiliation{Niels Bohr Institute, University of Copenhagen, Jagtvej 128, DK-2200, Copenhagen, Denmark}

\author[0000-0001-7160-3632]{Katherine E. Whitaker}
\affiliation{Department of Astronomy, University of Massachusetts Amherst, 710 N Pleasant Street, Amherst, MA 01003, USA}
\affiliation{Cosmic Dawn Center (DAWN), Copenhagen, Denmark}

\author[0000-0001-7095-7543]{Min Yun}
\affiliation{Department of Astronomy, University of Massachusetts Amherst, 710 N Pleasant Street, Amherst, MA 01003, USA}

\author[0000-0002-6184-9097]{Jaclyn B. Champagne}
\affiliation{Steward Observatory, University of Arizona, 933 N. Cherry Ave, Tucson, AZ 85719, USA}

\author[0000-0003-4761-2197]{Nicole E. Drakos}
\affiliation{Department of Astronomy and Astrophysics, University of California, Santa Cruz, 1156 High Street, Santa Cruz, CA 95064, USA}

\author[0000-0002-8008-9871]{Fabrizio Gentile}
\affiliation{University of Bologna - Department of Physics and Astronomy “Augusto Righi” (DIFA), Via Gobetti 93/2, I-40129 Bologna, Italy}
\affiliation{INAF- Osservatorio di Astrofisica e Scienza dello Spazio, Via Gobetti 93/3, I-40129, Bologna, Italy}

\author[0000-0001-9885-4589]{Steven Gillman}
\affiliation{Cosmic Dawn Center (DAWN), Copenhagen, Denmark}
\affiliation{DTU-Space, Technical University of Denmark, Elektrovej 327, DK-2800 Kgs. Lyngby, Denmark}

\author[0000-0002-0236-919X]{Ghassem Gozaliasl}
\affiliation{Department of Computer Science, Aalto University, PO Box 15400, Espoo, FI-00 076, Finland}
\affiliation{Department of Physics, Faculty of Science, University of Helsinki, 00014 Helsinki, Finland}

\author[0000-0002-7303-4397]{Olivier Ilbert}
\affiliation{Aix Marseille Univ, CNRS, CNES, LAM, Marseille, France  }

\author[0000-0002-8412-7951]{Shuowen Jin}
\affiliation{Cosmic Dawn Center (DAWN), Copenhagen, Denmark}
\affiliation{DTU-Space, Technical University of Denmark, Elektrovej 327, DK-2800 Kgs. Lyngby, Denmark}

\author[0000-0002-6610-2048]{Anton M. Koekemoer}
\affiliation{Space Telescope Science Institute, 3700 San Martin Dr., Baltimore, MD 21218, USA} 

\author[0000-0002-5588-9156]{Vasily Kokorev}
\affiliation{Kapteyn Astronomical Institute, University of Groningen, PO Box 800, 9700 AV Groningen, The Netherlands}

\author[0000-0001-9773-7479]{Daizhong Liu}
\affiliation{Max-Planck-Institut f\"ur Extraterrestrische Physik (MPE), Giessenbachstr. 1, D-85748 Garching, Germany}

\author[0000-0003-0427-8387]{R. Michael Rich}
\affiliation{Department of Physics and Astronomy, UCLA, PAB 430 Portola Plaza, Box 951547, Los Angeles, CA 90095-1547}

\author[0000-0002-4271-0364]{Brant E. Robertson}
\affiliation{Department of Astronomy and Astrophysics, University of California, Santa Cruz, 1156 High Street, Santa Cruz, CA 95064, USA}

\author[0000-0001-6477-4011]{Francesco Valentino}
\affiliation{Cosmic Dawn Center (DAWN), Copenhagen, Denmark}
\affiliation{Niels Bohr Institute, University of Copenhagen, Jagtvej 128, DK-2200, Copenhagen, Denmark}

\author[0000-0003-1614-196X]{John R. Weaver}
\affiliation{Department of Astronomy, University of Massachusetts, Amherst, MA 01003, USA}

\author[0000-0002-7051-1100]{Jorge A. Zavala}
\affiliation{National Astronomical Observatory of Japan, 2-21-1 Osawa, Mitaka, Tokyo 181-8588, Japan}

\author[0000-0001-9610-7950]{Natalie Allen}
\affiliation{Cosmic Dawn Center (DAWN), Copenhagen, Denmark}
\affiliation{Niels Bohr Institute, University of Copenhagen, Jagtvej 128, DK-2200, Copenhagen, Denmark}

\author[0000-0001-9187-3605]{Jeyhan S. Kartaltepe}
\affiliation{Laboratory for Multiwavelength Astrophysics, School of Physics and Astronomy, Rochester Institute of Technology, 84 Lomb Memorial Drive, Rochester, NY 14623, USA}

\author[0000-0002-9489-7765]{Henry Joy McCracken}
\affiliation{Institut d’Astrophysique de Paris, UMR 7095, CNRS, and Sorbonne Université, 98 bis boulevard Arago, F-75014 Paris, France}

\author[0000-0003-2397-0360]{Louise Paquereau} 
\affiliation{Institut d’Astrophysique de Paris, UMR 7095, CNRS, and Sorbonne Université, 98 bis boulevard Arago, F-75014 Paris, France}

\author[0000-0002-4485-8549]{Jason Rhodes}
\affiliation{Jet Propulsion Laboratory, California Institute of Technology, 4800 Oak Grove Drive, Pasadena, CA 91001, USA}

\author[0000-0002-7087-0701]{Marko Shuntov}
\affiliation{Institut d'Astrophysique de Paris, CNRS, Sorbonne Universit\'e, 98bis Boulevard Arago, 75014, Paris, France}

\author[0000-0003-3631-7176]{Sune Toft}
\affiliation{Cosmic Dawn Center (DAWN), Copenhagen, Denmark}
\affiliation{Niels Bohr Institute, University of Copenhagen, Jagtvej 128, DK-2200, Copenhagen, Denmark}

\begin{abstract}
A growing number of far-infrared bright sources completely invisible in deep extragalactic optical surveys hint at an elusive population of $z>4$ dusty, star-forming galaxies. Cycle 1 \textit{JWST} surveys are now detecting their rest-frame optical light, which provides key insight into their stellar properties and statistical constraints on the population as a whole. 
This work presents the \textit{JWST}/NIRCam counterpart from the COSMOS-Web survey to a far-infrared SCUBA-2 and ALMA source, AzTECC71, which was previously undetected at wavelengths shorter than 850\,\micron. AzTECC71, amongst the reddest galaxies in COSMOS-Web with ${\rm F277W - F444W}\sim0.9$, is undetected in NIRCam/F150W and F115W and fainter in F444W than other sub-millimeter galaxies identified in COSMOS-Web by $2-4$ magnitudes. This is consistent with the system having both a lower stellar mass and higher redshift than the median dusty, star-forming galaxy. With deep ground- and space-based upper limits combined with detections in F277W, F444W and the far-IR including ALMA Band 6, we find a high probability (99\%) that AzTECC71 is at $z>4$ with $z_{phot}=5.7^{+0.8}_{-0.7}$. This galaxy is massive ($\mathrm{\log\,M_*/M_\odot\sim10.7}$) and IR-luminous ($\mathrm{\log\,L_{IR}/L_\odot\sim12.7}$), comparable to other optically-undetected but far-IR bright dusty, star-forming galaxies at $z>4$. This population of luminous, infrared galaxies at $z>4$ is largely unconstrained but comprises an important bridge between the most extreme dust-obscured galaxies and more typical high-redshift star-forming galaxies.
If further far-IR-selected galaxies that drop out of the F150W filter in COSMOS-Web have redshifts $z>4$ like AzTECC71, then the volume density of such sources may be $\sim3-10\times$ greater than previously estimated. 
\end{abstract}

\section{Introduction\label{sec:intro}}
Infrared-bright galaxies have eluded optical detection since the first blind surveys at sub-millimeter (mm) wavelengths. Indeed, uncovering the optical counterpart to the first source ever discovered in an un-biased extragalactic survey at 850\micron, HDF850.1 \citep{Hughes1998}, spanned a 16-year long debate \citep[e.g.,][]{Dunlop2004,Cowie2009,Walter2012,Serjeant2014}. While deep optical/near-infrared imaging surveys have since significantly improved the detection rate of far-IR-bright galaxies, there is evidence that $\sim20-30\%$ of all sources in blind sub-mm surveys of extragalactic legacy fields have no counterpart at shorter wavelengths than \textit{Spitzer}/IRAC 3.6\micron\ \citep{Wardlow2011,Simpson2014,Casey2014,Franco2018}. 
The number densities of these sources and their contributions to the cosmic star-formation rate density at $z>3$ have been largely unconstrained, which may substantially alter our view of early star formation based on UV-bright Lyman-break galaxies \citep[e.g.,][]{MadauDickinson2014,Novak2017,Liu2018,Gruppioni2020,Zavala2021,Algera2023}.

The redshift distributions of optically-faint and far-infrared bright galaxies, particularly at $z>3$, represent a key uncertainty towards assessing the prevalance of dust-obscured star-formation at early times as well as the evolutionary role occupied by dusty galaxies. For instance, \cite{Long2022} show how recent $z\sim1-6$ number densities of dusty star-forming and quiescent galaxies broadly agree with the former as progenitors of the latter \citep[e.g.,][]{Toft2014,Valentino2020}.
However, as reviewed in \cite{Long2022}, dusty star-forming galaxy number counts above $z>4$ exhibit over 2 orders of magnitude in dispersion due to differences in area and wavelength coverage that produce drastically different levels of completeness in stellar mass, redshift, and volume; a more firm measurement on the prevalence of these sources at $z > 4$ will  constrain quenching timescales/mechanisms in the first 2 Gyr of the Universe \citep{Hayward2021}.
\cite{Casey2021} leverage the selection function of deep 2mm observations (RMS$\,=60-90\,\mu\rm{Jy\,beam^{-1}}$) to filter out most $z<3$ dusty star-forming galaxies, and find that directly detected 2mm-sources contribute 25\% of the star-formation rate density at $z\sim5$ and 10\% by $z\sim6$. 
Similarly, the ALPINE \citep{LeFevre2020,Bethermin2020,Faisst2020} and REBELS \citep{Bouwens2022} surveys find a large number of ``optically-dark'' sources at $z>4$ and $z>6.5$, respectively, suggesting their significant contribution to the star-formation rate density \citep{Gruppioni2020,Talia2021,Fudamoto2021}. 
At this epoch \cite{Barrufet2022} and \cite{Rodighiero2023} argue that a population of moderately obscured ($A_V\sim2$), red and very faint \textit{JWST} sources may contribute more to the star-formation rate density than previously thought (see also \citealt{PerezGonzalez2022}), but these samples lack the far-infrared data to directly constrain obscured star-formation. Naturally, a multi-wavelength approach to finding and measuring properties of dusty galaxies at $z>4$ is needed \citep[e.g.,][]{Zavala2022}.

While faint or undetected optical/near-IR photometry makes measuring stellar properties difficult in far-infrared bright galaxies, the limiting sub-mm spatial resolution poses an equal challenge in terms of counterpart identification. SCUBA-2 has been a powerful instrument for surveying wide fields in the sub-mm \citep*[e.g.,][]{Casey2014},  but its $\sim11^{\prime\prime}$ resolution ($\sim70$ kpc at $z=4-5$) at 850\micron\ allows for many possible optical counterparts within a given sub-mm source \citep[e.g.,][]{Liu2018,Jin2018}. ALMA and/or deep VLA imaging at high spatial resolution are critical for assigning far-IR emission to optical/near-IR counterparts \citep[e.g.,][]{Simpson2015,Zavala2018}. 
Working towards a better understanding of multi-wavelength counterpart selection in the era of large \textit{JWST} extragalactic fields coincident with sub-mm surveys, we present our analysis on the selection of optically-faint far-infrared bright galaxies and comment on the nature of one exceptional source at $z>4$ with no prior non-\textit{JWST} counterpart below $\lambda_{\rm obs}=850$\micron. Large area and deep \textit{JWST} surveys such as COSMOS-Web \citep{Casey2022} will have the potential to uncover many of these ``optically-faint'' sources and allow us to study their redshift distribution and physical properties.

\begin{figure*}
    \centering
    \includegraphics[width=\textwidth]{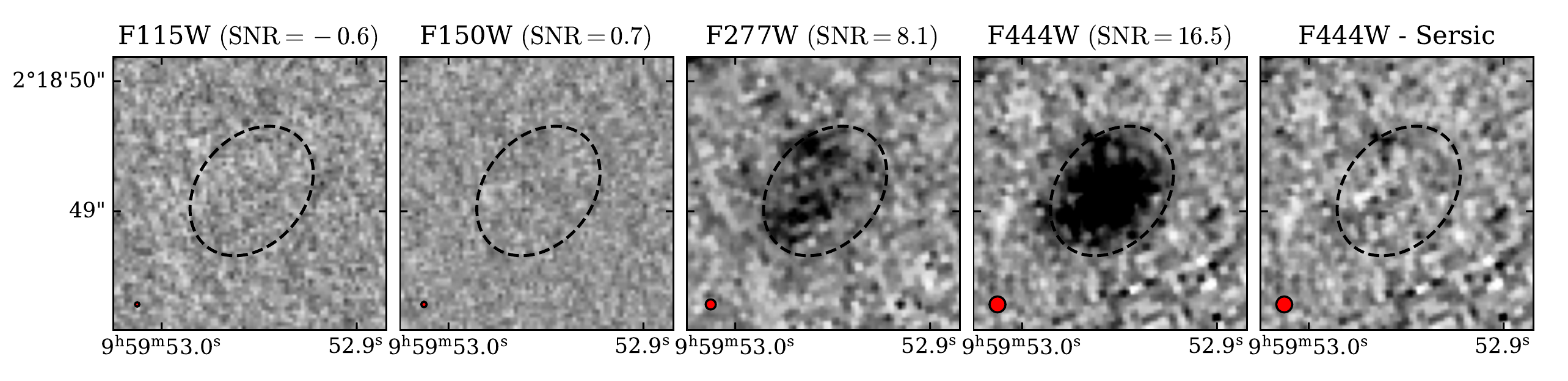}
    \caption{$2^{\prime\prime}\times2^{\prime\prime}$ \textit{JWST}/NIRCam cutouts from COSMOS-Web (Franco et al., in prep., \citealt{Casey2022}) of AzTECC71. On each panel we show the elliptical aperture fit to the F444W detection (black dashed ellipse). The right-most panel shows the residual between the F444W detection and a PSF-convolved 2D Sersic profile with $r_{1/2}=0.32^{\prime\prime}$ and $n_{\rm sersic}=0.74$. Red circles denote the PSF for each band shown. The signal-to-noise ratio of flux densities extracted through the elliptical aperture are listed above each panel. 
    }
    \label{fig:nircam_cutouts}
\end{figure*}

In this paper we present the \textit{JWST}/NIRCam detection of AzTECC71, a known far-infrared/sub-millimeter-only source \citep{Brisbin2017,Simpson2019}. We discuss the multi-wavelength associations and photometry in Section \ref{sec:data}, and spectral energy distribution fits to the data in Section \ref{sec:sedfitting}. In Section \ref{sec:disc} we discuss the nature of this source and implications for $4<z<6$ star-forming galaxy volume densities. Section \ref{sec:conclusion} summarizes the main conclusions. Throughout this work we adopt a $\Lambda$CDM cosmology with $\Omega_m=0.3$, $\Omega_\Lambda=0.7$ and $H_0=70\mathrm{\,km\,s^{-1}\,Mpc^{-1}}$. We use a Chabrier initial mass function (IMF).

\section{Data and Selection\label{sec:data}}

\subsection{Searching for near-infrared counterparts to sub-mm sources in COSMOS-Web}\label{sec:search}
Large area \textit{JWST} surveys are well-suited to uncovering the rest-frame optical emission originating from the intrinsically rare population of high-redshift dusty, star-forming galaxies that are far-IR/sub-mm bright \citep*{Casey2014,HodgeDaCunha2020}.
In this work, we search for ALMA counterparts to SCUBA-2 850\micron\ sources with signal-to-noise ratios (SNR) greater than 4 from S2COSMOS \citep{Simpson2019} within the COSMOS-Web January 2023 mosaic (Franco et al., in prep., \citealt{Casey2022}). 
Upon completion, COSMOS-Web will map a contiguous 0.54 deg$^2$ area within the COSMOS Survey \citep{Scoville2007} in four \textit{JWST}/NIRCam bands (F115W, F150W, F277W, F444W) and a non-contiguous 0.19 deg$^2$ area in MIRI/F770W. 
The January 2023 data includes six visits covering just 4\%\ ($77.76$ arcmin$^2$) of the total COSMOS-Web area. 

The data reduction of COSMOS-Web will be described in full in M. Franco et al., in prep. In summary, we reduce the \textit{JWST}/NIRCam data with the \textit{JWST} Calibration Pipeline version 1.8.3 with modifications for background and $1/f$ noise subtraction following other \textit{JWST} extragalactic programs \citep[e.g.,][]{Bagley2022,Finkelstein2022b}. We use version 0989 of the Calibration Reference Data System\footnote{\url{jwst-crds.stsci.edu}}.
The final mosaics have a resolution of $0.03^{\prime\prime}$/pixel, and have
been aligned to COSMOS2020 which in turn has been aligned to Gaia-EDR3 \citep{Weaver2022}. See \cite{Casey2022} for $5\sigma$ depths in each of the four COSMOS-Web NIRCam filters.  

Far-IR interferometric observations of the dust continuum emission are critical for a robust optical/near-IR counterpart identification to the low-resolution sub-mm SCUBA-2 data.
The \textit{JWST} counterparts and general properties of a larger ALMA/sub-mm selected sample from COSMOS-Web will be discussed in a future work (Manning et al.,~in prep). In this study we present the analysis of the faintest  NIRCam/F444W sub-mm source in our catalog, AzTECC71.

AzTECC71 was originally detected at 1.1 and 1.2 mm in the AzTEC/ASTE and MAMBO/IRAM 30m maps of the COSMOS field but its redshift was unknown due to lack of detections in the optical \citep{Aretxaga2011,Bertoldi2007}. It is detected at 850\micron\ by SCUBA-2 \citep{Simpson2019}, but not at 450\micron\ \citep[e.g.,][]{Casey2013,Geach2017,Lim2020}. \cite{Brisbin2017} conducted an ALMA Band 6 (1250\micron) follow-up survey of bright SCUBA-2 sources in COSMOS, and report a $6.3\sigma$ detection for AzTECC71 at RA, DEC = 9h59m52.95s, 2d18m49.13s from program 2013.1.00118.S (PI: M. Aravena), coincident with a NIRCam/F444W source in COSMOS-Web. 
Prior to the identification of its NIRCam/F444W counterpart AzTECC71 had no reported $>5\sigma$ detection below 850\micron\ including data from \textit{HST} and \textit{Spitzer}.  As there are no other optical/near-IR counterparts within 1$^{\prime\prime}$ of the ALMA detection, AzTECC71 is not in the COSMOS2020 catalog \citep{Weaver2022}. AzTECC71 will not be covered by COSMOS-Web's MIRI mosaic. 

Figure \ref{fig:nircam_cutouts} shows AzTECC71's \textit{JWST}/NIRCam counterpart from COSMOS-Web (Franco et al., in prep., \citealt{Casey2022}). 
The galaxy is detected in F277W and F444W but shows no detection in NIRCam's F115W and F150W bands with respective depths of 27.45 and 27.66 AB magnitudes. From a PSF-convolved 2D Sersic surface brightness profile fit to the F444W map (Fig.~\ref{fig:nircam_cutouts} \textit{right}), AzTECC71 has a half-light radius ($r_{1/2}$) of $0.32^{\prime\prime}\pm0.01^{\prime\prime}$ at 4.44\micron, and a Sersic index ($n_{\rm sersic}$) of $0.74\pm0.02$. Errors on $r_{1/2}$ and $n_{\rm sersic}$ are bootstrapped from 1000 perturbations of the F444W map with noise drawn from the background pixel flux distribution within 4$^{\prime\prime}$ of AzTECC71. We use the strong F444W detection to refine upper limits from the ground- and space-based imaging data, as described in the next section.

\subsection{Optical/Near-IR Photometry}

AzTECC71 shows an extended F444W morphology (Fig.~\ref{fig:nircam_cutouts}); Therefore, we measure photometry and upper limits using an elliptical aperture constructed to match the source morphology in F444W with a semi-major axis of $a=0.55''$, an axis ratio of $b/a=0.75$, and a position angle of $\phi = 40^\circ$ N of W. 
The semi-major/minor axes are greater than $r_{1/2}$ from the 2D Sersic fits and encase the extent of pixels $>5\sigma$ in the F444W map (Fig.~\ref{fig:nircam_cutouts}). We sum all pixels within the ellipse when calculating flux densities, and emphasize that this is not model-based photometry. AzTECC71 is detected at a SNR of $16.5$ in F444W ($m_{\rm F444W,AB}=24.62$) and $8.1$ in F277W ($m_{\rm F277W,AB}=25.51$). The galaxy is not detected in either of the other COSMOS-Web \textit{JWST}/NIRCam bands (F115W, F150W) nor any other optical/near-IR imaging in COSMOS. We calculate upper limits from these non-detections by summing the corresponding pixels within the elliptical aperture. Then we apply aperture corrections to account for PSF variations by computing the fraction of each lower resolution PSF that falls outside of our elliptical aperture shown on Figure \ref{fig:nircam_cutouts}. This is most relevant for the ground-based data in which the source would be unresolved. For the ground-based data, we adopt the PSFs used in the COSMOS2020 catalog \citep{Weaver2022}. The aperture corrections to lower-resolution ground-based imaging range from $2.1-2.4$, and is $1.1$ for \textit{HST/}ACS \citep{Koekemoer2007}. 
Additionally, using our refined aperture we recover a $3\sigma$ detection in IRAC(4.5\micron) where no detection was previously reported for lack of an optical/near-IR counterpart. 
\textit{JWST}/NIRCam and updated flux densities and limits are listed in Table \ref{tab:data}.

\begin{table}
	\centering
	\caption{AzTECC71 multi-wavelength photometry}
	\label{tab:data}
	\begin{tabular}{lccr} 
    \hline
    Band  & $\lambda_c$  & Unit & Flux \\
		\hline
CFHT-$u$ & 375 nm & nJy   & $(-14.1 \pm 9.1)$           \\
HSC-$g$ & 476 nm & nJy    & $(-2.9 \pm 10.2)$           \\
HSC-$r$ & 617 nm & nJy    & $(4.5 \pm 17.1)$            \\
HSC-$i$ & 768 nm & nJy   & $(13.1 \pm 29.6)$            \\
ACS/F814W & 814 nm & nJy & $(2.0 \pm 7.6)$              \\
HSC-$z$ & 891 nm & nJy   & $(28.1 \pm 26.8)$            \\
HSC-$y$ & 976 nm & nJy   & $(-18.0 \pm 55.2)$           \\
UVISTA Y  & 1.02~\micron & $\mu$Jy & $(0.11 \pm 0.11)$  \\
UVISTA J  & 1.25~\micron & $\mu$Jy & $(0.17 \pm 0.14)$  \\
UVISTA H  & 1.63~\micron & $\mu$Jy & $(0.14 \pm 0.16)$  \\
UVISTA Ks & 2.14~\micron & $\mu$Jy & $(0.16 \pm 0.12)$  \\
NIRCam/F115W & $1.15$~\micron & nJy & $(-31 \pm 51)$    \\
NIRCam/F150W & $1.50$~\micron & nJy& $(33 \pm 50)$      \\
NIRCam/F277W & $2.77$~\micron & nJy& $227 \pm 28$       \\
NIRCam/F444W & $4.44$~\micron & nJy& $513 \pm 31$       \\
IRAC/Ch1 & $3.6$~\micron & nJy & $(508 \pm 274)$        \\
IRAC/Ch2 & $4.5$~\micron & nJy & $706 \pm 228$          \\
IRAC/Ch3 & $5.8$~\micron & $\mu$Jy & $(1.6 \pm 0.30)$   \\
IRAC/Ch4 & $8.0$~\micron & $\mu$Jy & $(1.8 \pm 0.41$)   \\
MIPS\tablenotemark{a} & 24~\micron\ & $\mu$Jy & ($-0.22\pm18$)           \\
PACS\tablenotemark{b} & 100~\micron\ & mJy & $(-0.01\pm1.5)$             \\
PACS\tablenotemark{b} & 160~\micron\ & mJy & $(0.13\pm2.9)$              \\
SPIRE & 250~\micron\ & mJy& ($14.5\pm5.8$)              \\
SPIRE & 350~\micron\ & mJy& ($28.5\pm6.3)$              \\
SCUBA-2 & 450~\micron\ & mJy & $(4.0\pm6.1)$            \\
SPIRE & 500~\micron\ & mJy& ($27.2\pm6.1)$              \\
SCUBA-2 & 850~\micron\ & mJy & $4.31\pm0.78$            \\
AzTEC\tablenotemark{c} & 1100~\micron & mJy & $2.4\pm1.1$                \\
MAMBO\tablenotemark{d} & 1200~\micron & mJy & $3.1\pm1.3$                \\
ALMA/B6 & 1250~\micron & mJy & $2.16\pm0.20$            \\
ALMA/B4 & 2 mm & mJy & $(0.30\pm0.23)$                  \\
VLA/3 GHz & 10 cm & $\mu$Jy & ($3.2\pm2.3$)  \\ 
\hline
	\end{tabular}\\
 \tablecomments{ Flux densities reported in parenthesis correspond to limits on the SED as discussed in Section \ref{sec:data}. In general, we report pixel values at the ALMA/B6 position and their $1\sigma$ uncertainties. ($a$) \cite{LeFloch2009}. ($b$) \cite{Lutz2011}. ($c$) \cite{Aretxaga2011}. ($d$) \cite{Bertoldi2007}.}
\end{table}
\subsection{IR through radio data}
\begin{figure}
    \centering
    \includegraphics[width=.43\textwidth]{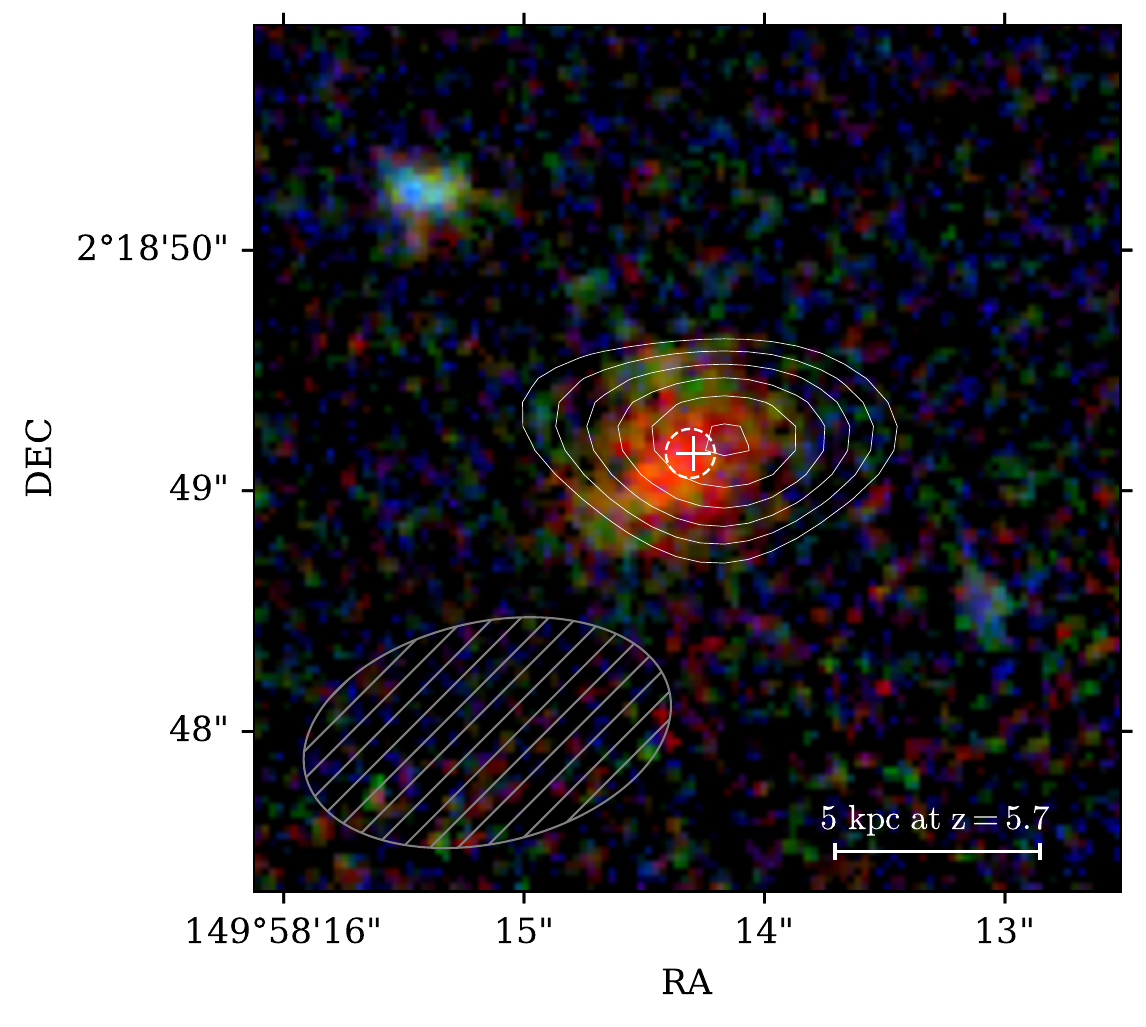}
    \caption{$3.6^{\prime\prime}\times3.6^{\prime\prime}$ RGB image constructed from R=F444W, G=F277W, B=(F115W+F150W stack). White contours are ALMA Band 6 continuum at $1250$\micron\ drawn at 5 through 10$\sigma$ in increments of $1\sigma$. The source is not spatially resolved by ALMA. The white circle corresponds to the centroid of the F444W image. The white $+$ is the centroid of a 2D Gaussian fit to the ALMA detection.
    }
    \label{fig:rgb}
\end{figure}

AzTECC71 is detected at: 850\micron\ (SCUBA-2, \citealt{Simpson2019}), 1100\micron\ (AzTEC/\citealt{Aretxaga2011}), 1200\micron\ (MAMBO/\citealt{Bertoldi2007}), and 1250\micron\ (ALMA B6/\citealt{Brisbin2017}) where the peak pixel flux within the elliptical aperture fit to the F444W map is $>5\times$ the map noise. In the case of the ALMA Band 6 detection, we restore AzTECC71's calibrated visibilities from Project 2013.1.00118.S (PI: Aravena) hosted in the ALMA archive. We image the data with \texttt{tclean} and naturally-weighted visibilities. The source is not spatially resolved by the $1.55^{\prime\prime}\times0.92^{\prime\prime}$ synthesized beam, so the naturally-weighted image maximizes the SNR. We then calculate the peak flux density from the primary-beam-corrected image, listed in Table \ref{tab:data}. 
We estimate the maximum size of the source in the Band 6 data from Equation 1 in \cite{MartiVidal2014}, which limits the size to $<0.44^{\prime\prime}$ (consistent with the F444W radius, see Section \ref{sec:search}). The ALMA Band 6 continuum detection is shown over an RGB image constructed from the NIRCam bands in Figure \ref{fig:rgb}.

\begin{figure}
    \centering
    \includegraphics[width=.47\textwidth]{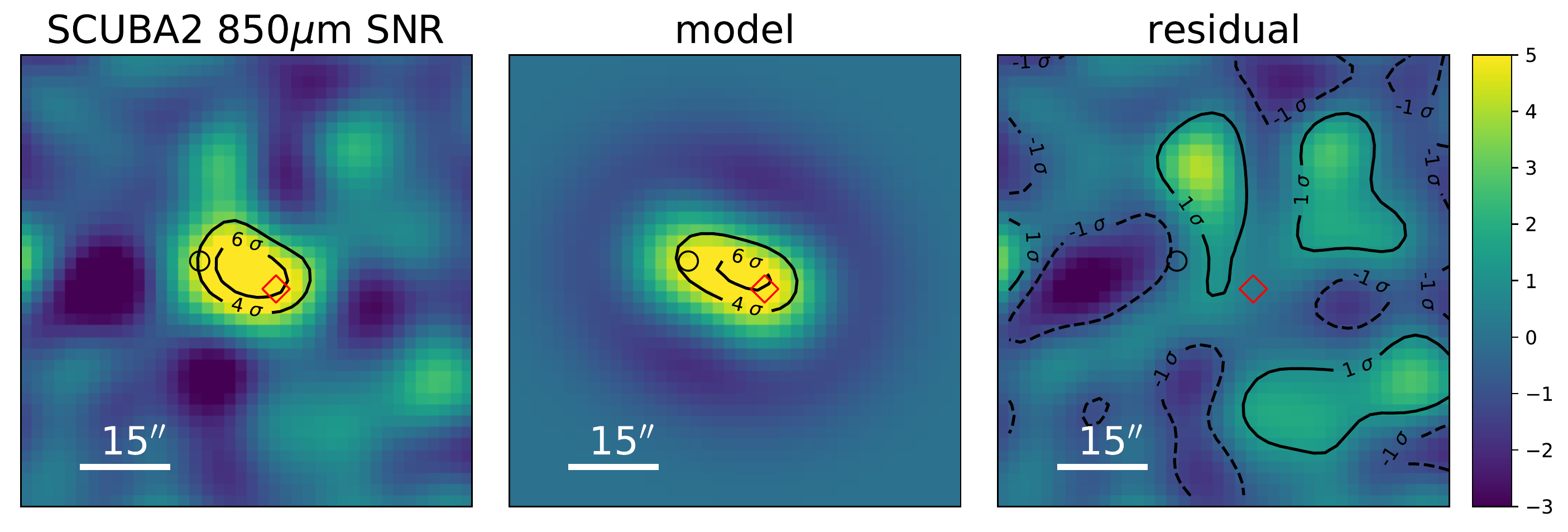}
    \caption{$80^{\prime\prime}\,\times\,80^{\prime\prime}$ cutout of the SCUBA-2 SNR map (\textit{Left}) illustrating the deblending of the SCUBA-2 850\micron\ flux using ALMA Band 6 positional priors for AzTECC71 (red diamond) and the nearby $z=0.829$ dusty galaxy (black circle). The model based on two scaled point sources fixed to the Band 6 positions is shown in the middle panel, with the residual shown in the far right. 
    }
    \label{fig:deblend}
\end{figure}

Two sub-mm sources are detected in the ALMA/B6 imaging of AzTECC71 at SNR$\,>5$, one coincident with the \textit{JWST}/NIRCam imaging and another 15$^{\prime\prime}$ away with a spectroscopic redshift of $z_s=0.829$ (AzTECC71b, \citealt{Brisbin2017}). Both sources contribute to blended sub-mm flux measured by the single-dish facilities, as is common for $10-20\%$ of all sub-millimeter galaixes (SMGs, $S_{850\mu m}>1$ mJy)  \citep{Chen2013,Koprowski2014,Michalowski2017,Hayward2013,Hayward2018}. The ALMA map provides secure positional priors for the origin of the sub-mm emission from each galaxy, so we deblend the SCUBA-2 flux densities by fitting point-source models fixed to the positions of each ALMA source.  We use the SCUBA-2 PSF from \cite{Simpson2017}. The results are shown in Figure \ref{fig:deblend} with the updated flux density for AzTECC71 listed in Table \ref{tab:data} at an SNR of $5.5$. The relative contribution of AzTECC71 to the total blended SCUBA-2 flux density is $55^{+1}_{-3}\%$. The FIR data at $\lambda_{obs}>850$\micron\ is beyond the SED peak for both AzTECC71 and the sub-mm source at $z=0.829$ so we scale AzTECC71's AzTEC ($34^{\prime\prime}$ resolution) and MAMBO ($11^{\prime\prime}$ resolution) fluxes by this same factor. We do not apply this to the blended \textit{Herschel} data which covers the SED peak of both sources, where differences in dust temperatures between the two will change the relative scaling band-to-band.

In the case of non-detections in the mid- and far-IR, we set the flux density to that of the pixel containing our target and adopt the RMS derived for the total mosaic as the $1\sigma$ uncertainty. In the case of \textit{Herschel}/SPIRE, this uncertainty corresponds to the confusion limit. We thereby constrain the flux/uncertainty at $24$\micron\ (\textit{Spitzer}/MIPS, \citealt{LeFloch2009}), $100-500$\micron\ (\textit{Herschel}/PACS+SPIRE, \citealt{Lutz2011,Oliver2012}), $450$\micron\ (SCUBA-2, \citealt{Casey2013}), 2mm (ALMA/B4, Long in prep., see also \citealt{Casey2021}), and 3 GHz (VLA, \citealt{Smolcic2017}) using the corresponding imaging data in the COSMOS field. For the \textit{Herschel}/SPIRE bands this gives the appearance of a detection (Tab.~\ref{tab:data}); however, the SPIRE data at the position of AzTECC71 is likely blended with neighboring sources as is common in confusion-limited SPIRE maps and as expected based on the sub-mm blending. Thus, the overlapping pixel contains some mix of emission from AzTECC71 and its neighbors. We test deblending the \textit{Herschel} maps following the method outlined previously for the SCUBA-2 data, but this is highly uncertain because of the larger \textit{Herschel} PSFs and higher confusion noise limits ($\sigma_{\mathrm{conf}}$). Our attempts at \textit{Herschel} deblending yields fluxes below the confusion limit for AzTECC71, and thus does not adequately constrain the source emission. Rather, by using the pixel flux at the position of the source $\pm\sigma_{\mathrm{conf}}$ we take a conservative approach and allow the SED fits to marginalize over uncertainty in flux association with AzTECC71 and its neighbors for the blended SPIRE photometry.

\begin{figure*}
    \centering
    \includegraphics[width=\textwidth]{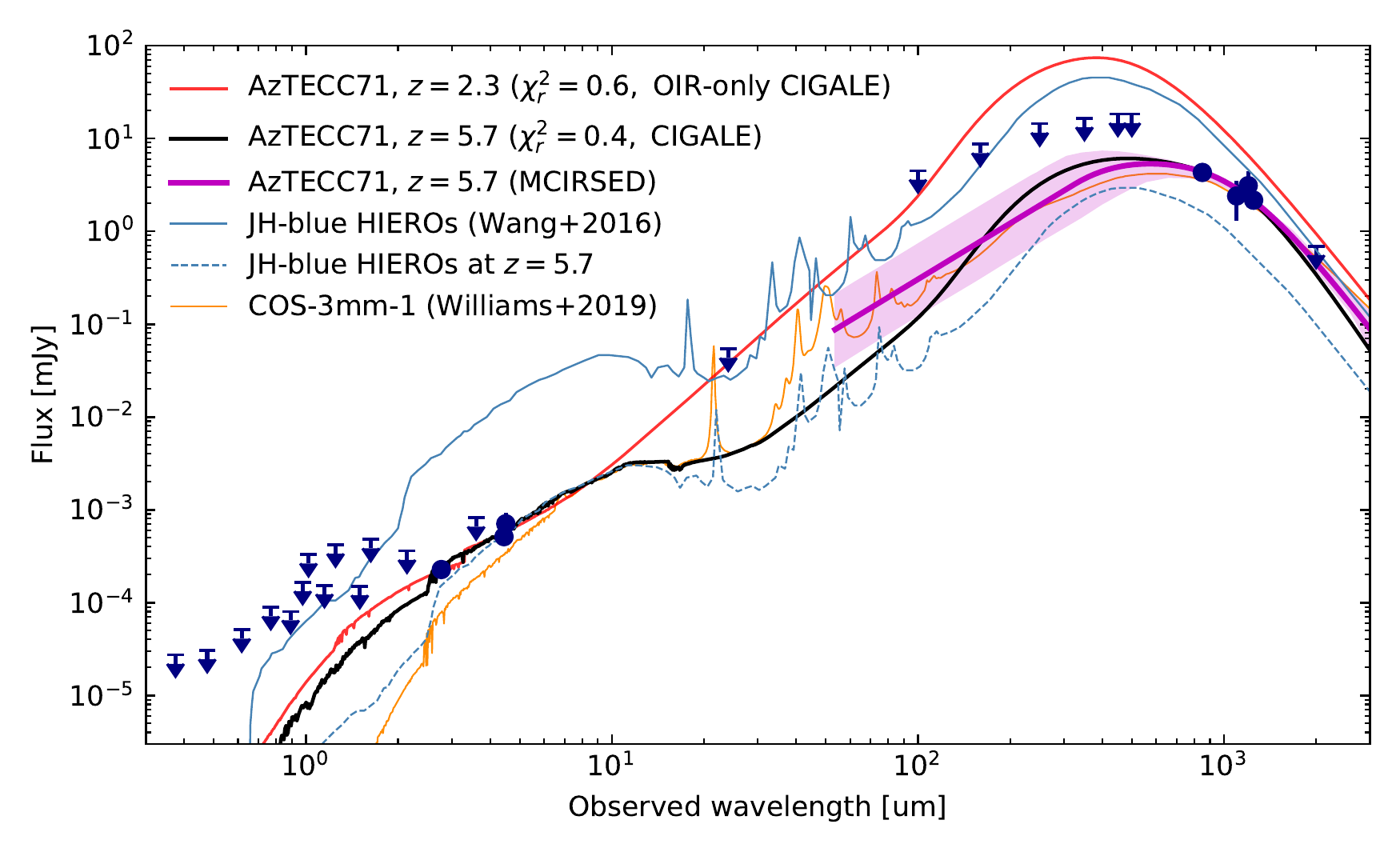}
    \caption{Spectral energy distribution for AzTECC71. Observations are shown in blue, with circles denoting detections and arrows $3\sigma$ upper limits. The preferred $z=5.7$ solution is shown for \texttt{CIGALE} with a black solid line, as well as a $z\sim2.3$ solution from fitting only the optical/near-IR SED (red line). The optical/near-IR only fit is largely ruled out by the far-IR/sub-mm data. The purple shaded region contains the uncertainty about the \texttt{MCIRSED} best-fit (solid purple line). For comparison we show the average SED of $z>3$ $H-[4.5]$ selected galaxies from \cite{Wang2016} in blue (solid). The best-fit SED to COS-3mm-1, a $z\sim5$ 3mm-selected galaxy from \cite{Williams2019} is shown in orange. AzTECC71 is fainter in the optical than $H-[4.5]$ selected massive and dusty star-forming galaxies from \cite{Wang2016}, but brighter than this sample in the far-IR if it were scaled to match AzTECC71 at $z=5.7$ and 4.5\micron\ (blue dashed). Similar to the IR flux of COS-3mm-1, AzTECC71 is likely part of a more IR-luminous population of massive galaxies at $z>4$.  
    }
    \label{fig:sed}
\end{figure*}

\section{SED Fits and Derived Properties\label{sec:sedfitting}}
We fit the optical through radio photometry/limits listed in Table \ref{tab:data} with \texttt{CIGALE} \--- a multi-wavelength  fitting code that handles UV/optical and IR energy balance \citep{Boquien2019}. We also fit the far-IR data only with: \texttt{MCIRSED} \--- a  Bayesian tool that fits dust emission properties \citep{Drew2022}, and \texttt{MMpz} \--- a photometric redshift code tied solely to far-IR/millimeter photometry \citep{Casey2020}. The FIR/mm probability distributions are based on the measured distribution of galaxy SEDs in the empirical relation between rest-frame peak wavelength and total IR luminosity, i.e. the $L_{\rm IR}$-$\lambda_{\rm peak}$ plane described in detail in \cite{Drew2022} and which does not evolve with redshift. This technique accounts for intrinsic SED breadth as it probes a wide range of dust temperatures at fixed IR luminosity. Based on the FIR through sub-mm constrains alone, we estimate $z_{p,\rm MMpz}=4.2^{+3.1}_{-1.6}$ from \texttt{MMpz}.

The full optical/near-IR to far-IR/sub-mm fit with  \texttt{CIGALE} is critical for constraining the target's redshift and stellar mass. We fit AzTECC71 with an exponentially declining star-formation history that allows for a late-stage burst with $\tau_{\rm main}=0.1,1$ Gyr and $\tau_{\rm burst}=1,10,100$ Myr.
We assume a Chabrier IMF \citep{Chabrier2003}, a metallicity of either $Z_\odot$ or $0.2Z_\odot$ and a power-law dust attenuation curve $\propto\lambda^{-0.7}$ up to $A_V=6$. We model the FIR SED as a modified blackbody with sub-mm slope $\beta\in[1.8,3.2]$ added to a mid-infrared power-law with a slope $\alpha\in[1,5]$ that accounts for a distribution in warmer dust temperatures \citep{Casey2012}, although this regime of the SED is largely unconstrained so $\alpha$ is a nuisance parameter we marginalize over.
We apply a flat prior on redshift between $z=2-9$, and the dust temperature is allowed to vary between 20 and 70 K. The wavelength corresponding to an optical depth of unity ($\lambda_0$) is fixed to 200\micron; we account for different opacity models in subsequent fits and find this assumption to have little impact on the best-fit results besides inflating uncertainties when allowed to vary. 
We include a power-law synchrotron component constrained by our VLA/3 GHz upper limit with a slope of 0.8 and we assume an FIR/radio correlation coefficient $q_{\rm IR}\in[1.8,2.6]$, corresponding to the range found for massive and high-redshift star-forming galaxies \citep[e.g.,][]{Delvecchio2021}. 

With \texttt{MCIRSED} \citep{Drew2022}, we fit the FIR/millimeter SED to a modified blackbody added piecewise with a mid-IR power law using Bayesian analysis; best-fit SEDs are derived based on a Markov chain Monte Carlo convergence. The mid-IR power law is joined to the modified blackbody at the point where the blackbody slope is equal to the power-law index $\alpha_{\mathrm{MIR}} = 2$ \citep[consistent with e.g.,][]{Casey2012,U2012}. We allow $\lambda_0$ to vary between $\lambda_{\mathrm rest} = 100-300$\,\micron\ as the opacity model for high-redshift galaxies likely varies as a function of the dust geometry \citep[e.g.,][]{Simpson2017,Jin2019,Jin2022a}. We fix the redshift to the \texttt{CIGALE} best-fit photometric redshift ($z=5.7$) and input the FIR/millimeter photometric detections and upper limits with their associated uncertainties. Given our prediction that the galaxy sits at $z\gtrsim5$, we include a cosmic microwave background (CMB) correction
term in our fitting procedure to account for ISM dust heating
from the CMB at high redshift \citep{daCunha2013}. From the \texttt{MCIRSED} algorithm we find the best-fit dust SED with measurements for each of the following free parameters:
emissivity spectral index ($\beta$), total IR luminosity (\lir, taken from 8 to 1000\,\micron), dust temperature ($\mathrm T_{\mathrm {dust}}$), and rest-frame peak wavelength ($\lambda_{\mathrm {peak}}$). We marginalize over $\lambda_0$ which increases the uncertainties on $\lambda_{\rm peak},\,\mathrm{T_{dust}}$ by 20\% compared to fits fixing the opacity model to $\lambda_0 = 200$\,\micron. 

\begin{figure}
    \centering
    \includegraphics[width=.49\textwidth]{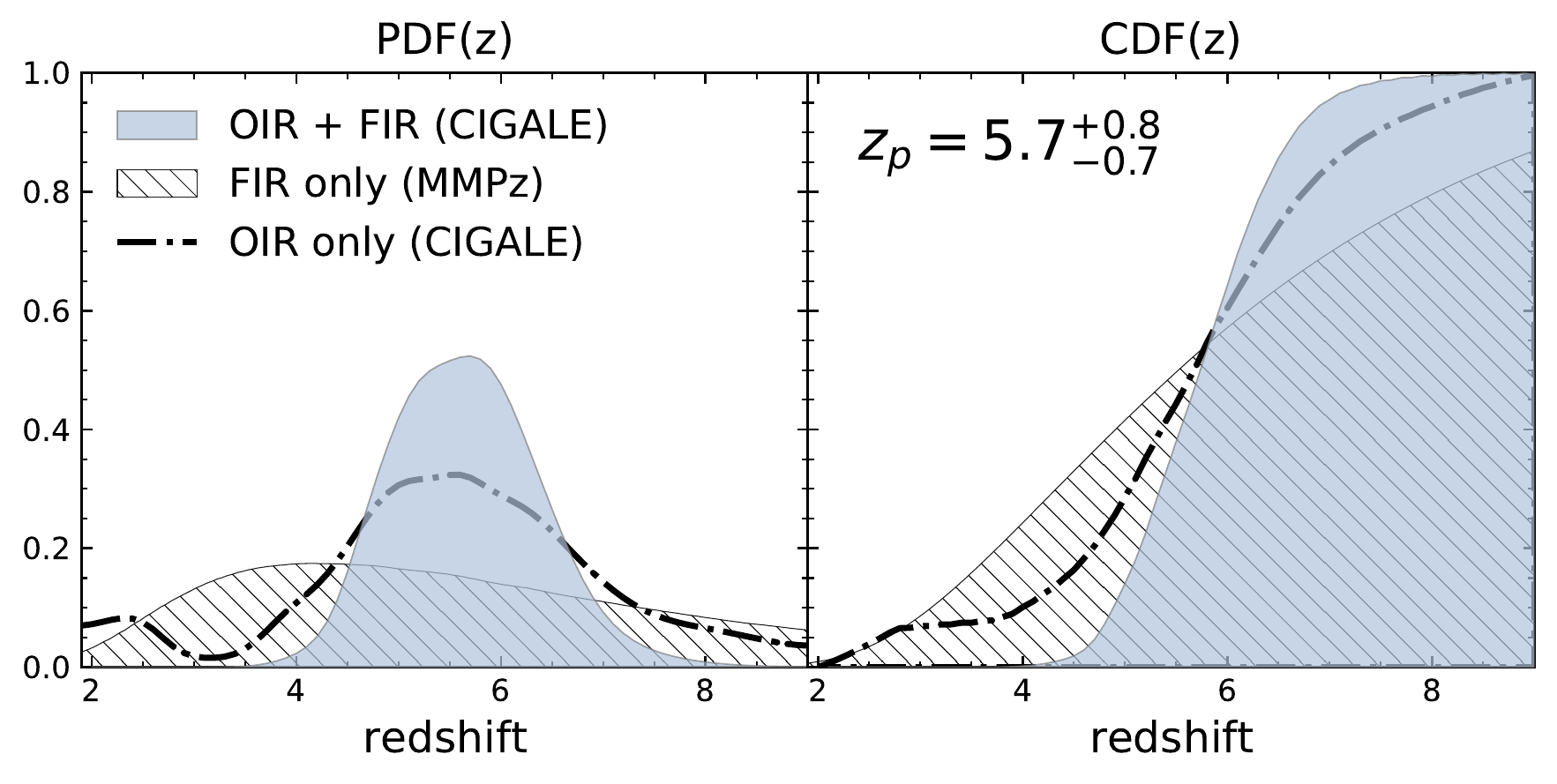}
    \caption{Posterior (\textit{left}) and cumulative (\textit{right}) redshift distribution function from fitting only the IR SED with \texttt{MMPz} (black hatched), only the optical/near-IR with \texttt{CIGALE} (dot-dashed), and the full optical through IR SED with \texttt{CIGALE} (blue shaded). We report the photometric redshift ($z_p$) from the peak of the optical/near-IR $+$ far-IR posterior distribution function and its 16th and 84th percentiles. 
    }
    \label{fig:pz}
\end{figure}

From the SED fits we infer a photometric redshift of $z_{p}=5.7^{+0.8}_{-0.7}$. 
Figure \ref{fig:sed} shows the best-fit \texttt{CIGALE} SED at $z=5.7\,\,(\chi_r^2=0.4)$. 
The posterior and cumulative redshift distribution functions are shown in Figure \ref{fig:pz} from both \texttt{CIGALE} and \texttt{MMPz}. Marginalizing over the full parameter space, the probability that the galaxy lies above $z=4$ ($z=5$) is 99.6\% (79.0\%). If we fit only the optical/near-IR (OIR) SED under the same assumptions as outlined previously, we find a much broader redshift posterior consistent with the OIR+FIR/sub-mm photo$-z$ albeit with greater uncertainty: $z_{p,\rm OIR}\sim5.7^{+1.3}_{-1.2}$. 
Most notably, the OIR-only SED fit allows a solution at $z\sim2.3$ that is significantly disfavored by the far-IR/sub-mm/radio data as demonstrated on Figure \ref{fig:sed}. 

In Table \ref{tab:source} we list the best parameter estimates and their uncertainties for the $z_p=5.7$ solution from the optical/near-IR $+$ far-IR/sub-mm/radio SED fits with parameter uncertainties marginalizing over the $\chi^2$ distribution. Based on the \texttt{CIGALE} and \texttt{MCIRSED} fits, AzTECC71 is most likely a massive ($\mathrm{\log\,M_*/M_\odot\sim10.7}$) dust-obscured star-forming galaxy with $\mathrm{\log\,L_{IR}/L_\odot\sim12.75}$ at $z\sim5.7$. This is consistent with the OIR-only SED fit which favors a massive ($\mathrm{\log\,M_*/M_\odot\sim10.6}$), optically attenuated ($A_V\sim3$) galaxy at $z\sim5.7^{+1.3}_{-1.2}$.
AzTECC71 must be heavily attenuated in the optical ($A_V\sim5$) to match the red F277W/F444W color and non-detections in the NIRCam short-wavelength bands. This can be interpreted with a recent dust-obscured starburst as is known for ultraluminous IR galaxies (ULIRGs) with \lir $\,\geq10^{12}$\,\lsun. 
Indeed, the dust-obscured star-formation rate is high at  $\sim1000\,\mathrm{M_\odot\,yr^{-1}}$ as implied by the total IR luminosity following \cite{Kennicutt1998}\footnote{$\mathrm{SFR_{IR}/[M_\odot\,yr^{-1}]\,=\,1.8\times10^{-10}\,L_{IR}/L_\odot}$}.

\begin{figure}
    \centering
    \includegraphics[width=.48\textwidth]{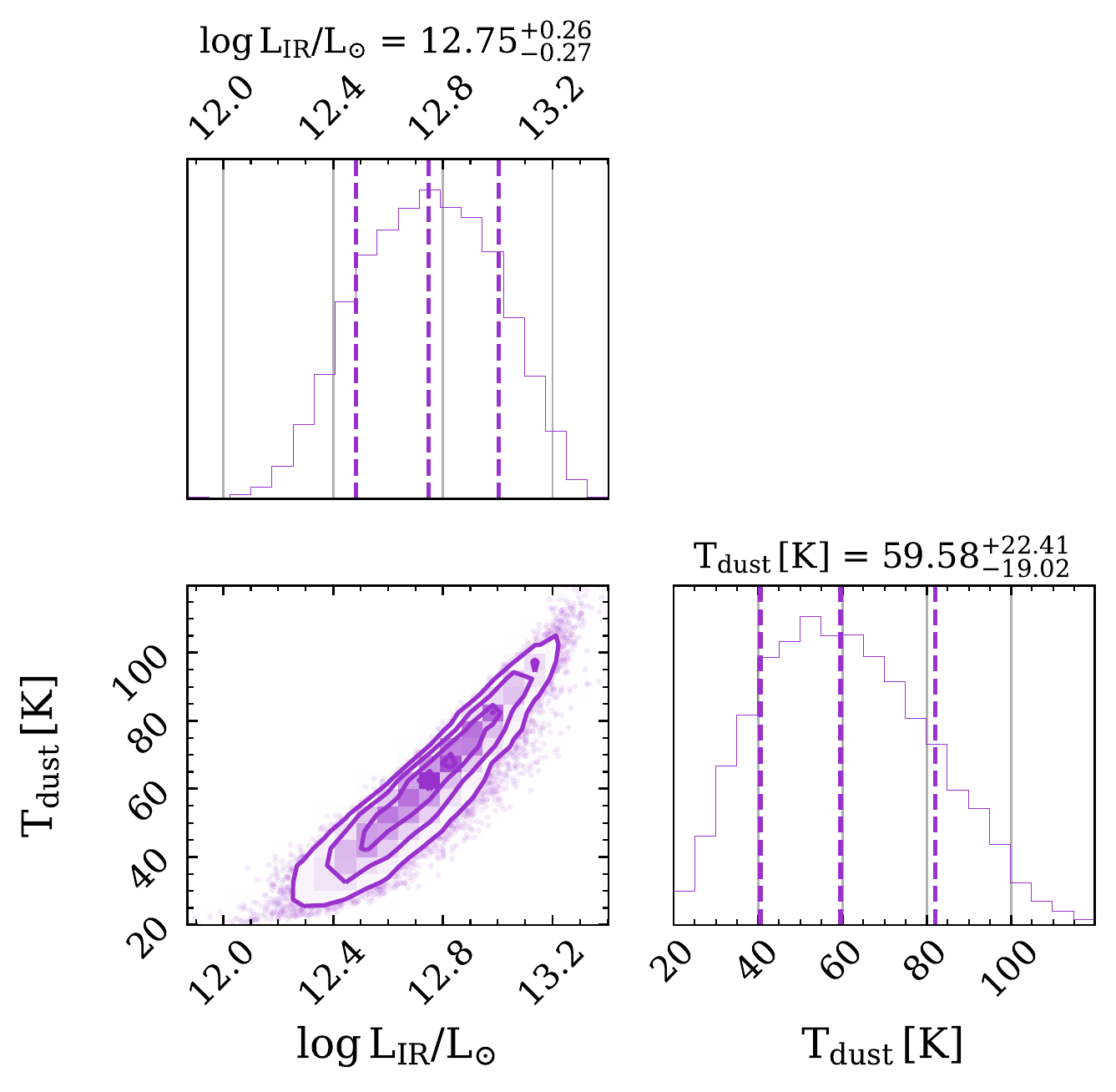}
    \caption{Corner plot showing the two-dimensional joint posterior distributions for dust SED fit parameters \lir\, and $\mathrm{T_{dust}}$ derived using \texttt{MCIRSED}. The vertical dashed lines in the 1D histograms for each parameters show the 68\% inter-quartile range (outer lines) and the median value (central line). The dust temperature and IR luminosity are highly covariant given the lack of rest-frame constraint straddling the dust SED peak. 
    }
    \label{fig:mcirsed}
\end{figure}

The \texttt{MCIRSED} dust SED fit favors $\lambda_{\mathrm {peak}}\sim80$\micron, corresponding to a dust temperature of $\sim60\pm20\,$K (see Figure \ref{fig:mcirsed}). 
While \texttt{MCIRSED} also fits the emissivity spectral index $\beta$, this measurement is poorly constrained, and is further exacerbated by the combined use of interferometric data with single-dish data that suffers from confusion boosting. Though we have accounted for deboosting and deblending as best as possible, precise measurements of $\beta$ for an individual source necessitates matched-beam ALMA data at both frequencies. Taken at face value, we find $\beta=2.7^{+0.6}_{-0.7}$, which is high compared to other estimates at $z\sim5.5$ \citep[e.g.,][]{Faisst2020tdust} but consistent with recent works that find evidence for $\beta\sim2.4$ in $z>4$ dusty, star-forming galaxies \citep{Cooper2022,Casey2021,Kato2018}.

Diagnosing the presence of an active galactic nucleus (AGN) in AzTECC71 is difficult given the lack of SED constraint in the mid-IR, a regime particularly sensitive to hot, toroidal dust emission around central super-massive black holes. 
At $z=5.7$ and for the best-fit \loglir$\,=12.75$, our radio upper limits at 3 GHz disfavor $q_{\rm IR}<2$ typical of radio AGN (\citealt[e.g.,][]{Delvecchio2021}) which would otherwise require a 3 GHz flux $3\times$ greater than our $3\sigma$ upper limit. However, a mid-IR spectral search for high-ionization emission lines and/or fine sampling of the mid-IR SED with \textit{JWST}/MIRI would be needed to robustly rule out the presence of a heavily obscured AGN.  

\subsection{Dust Mass}

We estimate the total dust mass in AzTECC71 following the procedure outlined in \cite{Kirkpatrick2017} using: 
\begin{equation}
    M\mathrm{_{dust}}=\frac{S_\nu D_L^2}{\kappa_\nu B_\nu (T_{\rm dust})}
    \label{eq:mdust}
\end{equation}
where $D_L$ is the luminosity distance at $z=5.7$, $S_\nu$ is the flux density, $B_\nu$ is the Planck equation, and $\kappa_\nu$ is the dust opacity from \cite{WeingartnerDraine2001} assuming MW-like dust and $R_V=3.1$. 
As explained in \cite{Kirkpatrick2017}, variation in $\kappa_\nu$ along the Rayleigh-Jeans (RJ) tail of the cold dust emission is $<10\%$ between MW, SMC, and LMC opacity models. This is lower than the measurement and model uncertainty limiting our dust mass calculation. 
We fix the cold dust temperature in our calculations to $T_{\rm dust}=25$ K because this is more representative of the mass-weighted dust temperature rather than the light-weighted dust temperature we get from the SED fits \citep{Scoville2016}. 

We use the \texttt{MCIRSED} best-fit SED 
flux and uncertainty at $\lambda_{\rm obs} =2$mm to calculate a dust mass using Equation \ref{eq:mdust}. At $z=5.7$, the 2mm flux is constrained by the ExMORA (Long et al., in prep.) upper-limit, and critically probes the RJ tail of cold dust emission ($\lambda_{\rm rest}\sim300\,$\micron). This regime is well-suited to measuring the total dust mass because (1) the temperature dependence along the RJ tail is linear, and (2) the emission is optically thin at long wavelengths in the sub-mm \citep{Scoville2014}. While there might be variation in the dust opacity law at $z>4$ \citep[e.g.,][]{Cooper2022},  \cite{Faisst2020tdust} find optically thin dust at $\lambda_{rest}>200$\micron\ in $z\sim5.5$ galaxies based on three ALMA bands sampling beyond the peak of the IR SED in UV-selected main-sequence galaxies.  

We then calculate $M_{\rm dust}$ by randomly sampling the range in $S_{\nu,\rm 2mm}$ from the \texttt{MCIRSED} fit (see purple shaded region, Fig.~\ref{fig:sed}), as well as the posteriors for 
$\beta$. We repeat this process $1000$ times, and take the most frequent $M_{\rm dust}$ as our estimate. We report upper and lower uncertainties from the 16th and 84th percentiles of the distribution. From these calculations we estimate the dust mass in AzTECC71 to be $\log M_{\rm dust}/M_\odot=8.1^{+0.3}_{-0.3}$. If we instead use the ALMA/B6 flux at $\lambda_{\rm obs}=1250$\micron\ following the same procedure, we get a value consistent within $1\sigma$ of $\log M_{\rm dust}/M_\odot=8.3^{+0.4}_{-0.2}$; however, 1250\micron\ may or may not be tracing the RJ tail given our photometric redshift uncertainties so we adopt the estimate anchored to the 2mm upper limit/SED as our fiducial dust mass. This value is close to the total dust mass amongst $z=0$ LIRGs, and $\sim1$ dex below the typical dust mass for $z\sim2$ LIRGs of similar stellar mass to AzTECC71 \citep{Kirkpatrick2017}. Applying this same method to the sample of \cite{Faisst2020tdust} to eliminate uncertainty introduced from using different methods to calculate the dust mass, we find a range in dust mass between $\log M_{\rm dust}/M_\odot=7.6-7.8\,(\pm0.3)$. Relative to these $z\sim5.5$ main-sequence galaxies from the UV-selected ALPINE survey \citep{Faisst2020tdust}, AzTECC71 has a greater dust mass by a factor of $\sim3$ and higher star-formation rate by a factor of $\sim8$ on-average. 

\begin{table}
	\centering
	\caption{AzTECC71 Derived Properties}
	\label{tab:source}
	\begin{tabular}{llll} 
		\hline
    Band  & Parameter  & Measurement & Units \\
		\hline
      All & $z_{phot}$ & $5.1_{-0.1}^{+0.8}$  & \nodata\\ 
      FIR & $z_{phot}$ & $4.2^{+3.1}_{-1.6}$ & \nodata \\ 
      All & $p(z>5)$ &  0.79 &\nodata \\ 
      All & $p(z>4)$ & 0.99 &\nodata \\
      All & $p(z<3)$ & $3\times10^{-5}$ & \nodata\\
      FIR & $p(z>4)$ & 0.74 & \nodata\\ 
            \hline 
    \multicolumn{2}{c}{Assuming $z=5.7$}&\\
    All & $\mathrm{M_*}$     & $7\pm3$ & $10^{10} \,\mathrm{M_\odot}$ \\
    All & $\mathrm{L_{IR}}$ & $4\pm3$ & $10^{12}\, \mathrm{L_\odot}$ \\
    All & $A_V$ & $5\pm1$ & \nodata \\ 
    FIR & $\mathrm{L_{IR}}$ & $6^{+4}_{-3}$ &  $10^{12}\, \mathrm{L_\odot}$ \\ 
    FIR & $\lambda_{\rm peak}$ & $83^{+25}_{-21}$ & \micron \\
    FIR & $\mathrm{T_{dust}}$ & $60^{+22}_{-19}$ & K \\
    FIR   & $\mathrm{\log\,M_{dust}}$ & $8.1^{+0.3}_{-0.3}$ & $\mathrm{M_\odot}$ \\
    FIR & $\mathrm{SFR_{IR}}$ & $1000^{+600}_{-500}$ & $\mathrm{M_\odot}$\,yr$^{-1}$ \\
    FIR & $\tau_{\mathrm{gas\,depl.}}$ & $12\pm8$ & Myr \\
    \hline	
     F444W & $n_{\rm sersic}$ & $0.74\pm0.02$ & \nodata \\
     F444W & $r_{1/2}$ & $0.32\pm0.01$ & arcsec \\ 
     B6   & $r_{\rm eff}$ & $<0.44$ & arcsec \\
     B6 & $\mathrm{\Sigma_{IR}}$ &  $>1.2$ & $10^{11}\,\mathrm{L_\odot\,kpc^{-2}}$\\
 \hline	
	\end{tabular}\\
\end{table}

\section{Results and Discussion \label{sec:disc}}

\subsection{Comparison to high-redshift samples of dusty, star-forming galaxies}
AzTECC71 stands out from high-redshift dusty, star-forming galaxies for its relatively faint optical and near-IR photometry. This galaxy is therefore closer in terms of observed flux densities to known ``optical/near-IR dark" (hereafter OIR-dark) samples selected in the far-IR/sub-mm \citep{daCunha2015,Williams2019,Manning2022}. Approximately $30\%$ of sub-mm-selected galaxies have always been unconstrainable in the optical/near-IR \citep{Wardlow2011,Simpson2014,Casey2014,Zavala2018}. \cite{Manning2022} report two 2mm-selected OIR-dark $z>3$ dusty galaxies, 
both of which are similar to AzTECC71 in terms of stellar mass and \lir. \cite{Williams2019} find a similar OIR-dark 3mm source likely at $z=5.5$, comparable to HDF850.1 after accounting for magnification \citep{Serjeant2014} and also similar to MAMBO-9 \citep{Casey2019,Jin2019}. 
\cite{Zavala2022} identify CEERS-DSFG1, an OIR-dark galaxy with a $2.25$ mJy SCUBA-2 $850\,\mu$m flux at $z=4.91$ \citep{Arrabal2023} detected only in F277W and longer filters like AzTECC71. CEERS-DSFG1 is less extreme with $\log\mathrm{L_{IR}/L_\odot}\sim12$ and $\log\mathrm{M_{*}/M_\odot}\sim10.3$ \citep{Zavala2022}, highlighting the population diversity within OIR-dark, far-IR bright sources. 
The properties of these sub-mm/mm-selected galaxies are consistent with radio-selected OIR-dark galaxies from \cite{Talia2021} and near-IR-faint SMGs from \cite{Smail2021}. That AzTECC71 falls within this OIR-dark, far-IR luminous $z\sim4-6$ dusty galaxy population is supported by similar SCUBA-2 flux densities and $3-5\sigma$ IRAC detections. 

AzTECC71 exhibits intrinsic stellar and dust properties common amongst larger SMG samples \citep[e.g.,][]{Simpson2019,Smail2021}. 
AzTECC71 has a dust-obscured star-formation rate of $1000\,M_\odot\,\mathrm{yr}^{-1}$ and stellar mass $7\times10^{10}\,\mathrm{M_\odot}$, corresponding to a specific star-formation rate (sSFR$\,\equiv\mathrm{SFR/M_*}$) of $14\pm\,\mathrm{Gyr^{-1}}$. This is $4\times$ greater than the sSFR of main-sequence galaxies at this epoch \citep{Speagle2014}. AzTECC71 also has a higher sSFR than IR-detected high-redshift galaxies: the average SFR and stellar mass of $z>2.5$ SMGs from the ALESS survey are $400\,\mathrm{M_\odot\,yr^{-1}}$ and $10^{11}\,\mathrm{M_\odot}$ respectively for a sSFR of 4 Gyr$^{-1}$ \citep{daCunha2015}. AzTECC71 is more massive than UV-selected $z>4$ star-forming galaxies with far-IR detections from ALMA-REBELS \citep{Inami2022} and ALPINE \citep{Bethermin2020,Faisst2020}, and has a larger dust mass by a factor of $\sim3$. This is consistent with AzTECC71 being a $z\sim5$ starburst possibly fueled by a large gas reservoir. Assuming a gas-to-dust ratio of 100 AzTECC71's depletion timescale ($\tau_{\mathrm{gas\,depl.}}\equiv M_{\rm gas}/\mathrm{SFR}$) is $\sim10$ Myr, significantly higher than that of $z\sim0-2$ LIRGs under identical dust mass assumptions \citep{Kirkpatrick2017}. Assuming no further gas accretion, AzTECC71 would deplete 99\% of its gas reservoir in $\sim50-100$ Myr and could therefore plausibly evolve into the emergent population of quiescent galaxies at $z\sim4-5$ \citep[e.g.,][]{Merlin2019,Santini2019,Shahidi2020,Long2022} 

Prior to \textit{JWST} AzTECC71 would not have been identified by rest-frame optical/near-IR methods for selecting high-$z$ dusty galaxies for lack of a  detection shortward of the sub-mm. In Figure \ref{fig:sed} we compare AzTECC71 against the average SED of $z>3$ \textit{HST} and \textit{Spitzer} $H-[4.5]$ selected objects from \cite{Wang2016} \--- ``HIEROs''. While thought to include a significant fraction of $z>3$ dusty star-forming galaxies, such HIEROs 
are $\sim$2 dex brighter in the near-IR than AzTECC71. HIEROs are much fainter in the far-IR/sub-mm than AzTECC71 if we normalize them to AzTECC71's redshift and IRAC(4.5\micron) flux, which suggests that AzTECC71 is not drawn from this OIR-faint galaxy population that falls between $3<z<6$ \citep{Wang2016}. In fact, AzTECC71 is even missed by H-dropout selection of OIR-dark IRAC sources despite having comparable $\sim850$\micron\ flux densities \citep{Wang2019}. This highlights the importance of both near- and far-IR/sub-mm selected samples of high$-z$ dusty, star-forming galaxies for completeness.  

\subsection{Morphology\label{sec:morphology}}

Spatially resolved optical and infrared emission in high-redshift, dusty galaxies commonly show offsets from one-another on the order of $0.2^{\prime\prime}-0.6^{\prime\prime}$ \citep[e.g.,][]{Franco2018,Elbaz2018}. This can arise from differential dust-attenuation across the galaxy, in particular due to clumpy dust distributions \citep[e.g.,][]{Seibert2005,Cortese2006,Boquien2009,MunozMateos2009,Faisst2017}. Interestingly, the near-infrared and 1250\micron\ continuum in AzTECC71 are remarkably coincident. Both the ALMA Band 6 peak and centroid  agree with the F444W centroid within $0.1$ arcsec (Fig.~\ref{fig:rgb}). While AzTECC71 is not spatially resolved by ALMA, the effective radius at 1250\micron\ must be below 0.44$^{\prime\prime}$ ($<2.6$ kpc at $z=5.7$), which could cover a large fraction of the stellar light given AzTECC71's half-light radius of $0.32^{\prime\prime}$ at 4.44\micron. AzTECC71 has a $r_{1/2\,\rm F444W}$ consistent with the range of near-IR sizes in $z\sim4$ $JH-$blue HIEROs from \cite{Wang2016}, and smaller than the average $H-$band sizes of $z\sim2$ SMGs from \cite{Swinbank2010} by $\sim40\%$. 
A high dust covering fraction could help explain the high $A_V=5$ needed to fit the rest-frame optical photometry, but there is also evidence from the RGB image (Fig.~\ref{fig:rgb}) that the galaxy is bluer towards the outskirts.
Indeed the F277W emission is clumpy and brighter away from the F444W and ALMA centroids, which suggests strong central attenuation.  
Spatially resolved far-IR observations are needed to fully test the resolved impact of dust on reddening across this galaxy only 1.2 Gyr after the Big Bang. Nevertheless, AzTECC71 likely hosts a very dusty nuclear starburst.

Given the size constraint from the ALMA Band 6 continuum detection, AzTECC71 must have a high IR surface density ($\Sigma\mathrm{_{IR}\equiv 0.5L_{IR}/\pi}r_{\rm eff}^2$) above $10^{11}\,\mathrm{L_\odot\,kpc^{-2}}$. This is consistent with the high IR surface densities observed for dusty, star-forming galaxies locally and at high-redshift \citep{DiazSantos2017,Fujimoto2017,Simpson2017,Jin2022}, and could help explain the warm dust temperature preferred by our FIR modelling as a compact starburst heats dust to high temperatures, more-so if the gas-phase metallicity is low \citep{RemyRuyer2014,Sommovigo2022}. Using radiative transfer modeling \cite{Hirashita2022} argue that dust temperatures of $\sim40$ K and above could possibly be explained with lower dust-to-gas ratios. Temperatures below 40 K are not preferred by our far-IR SED modelling (Fig.~\ref{fig:mcirsed}), but coverage over the far-IR SED peak is needed to robustly measure $\mathrm{T_{dust}}$ given degenerative solutions with both \lir\ and $\beta$. 
Taking these measurements at face value suggests a slightly higher dust temperature for this dust-obscured galaxy, which could be related to lower dust-to-gas ratios, metallicity, and/or a heavily obscured AGN which we cannot rule out for lack of data in the rest-frame mid-IR \citep[e.g.,][]{Kirkpatrick2015,McKinney2021agn}. 

\begin{figure*}[h!]
    \centering
    \includegraphics[width=\textwidth]{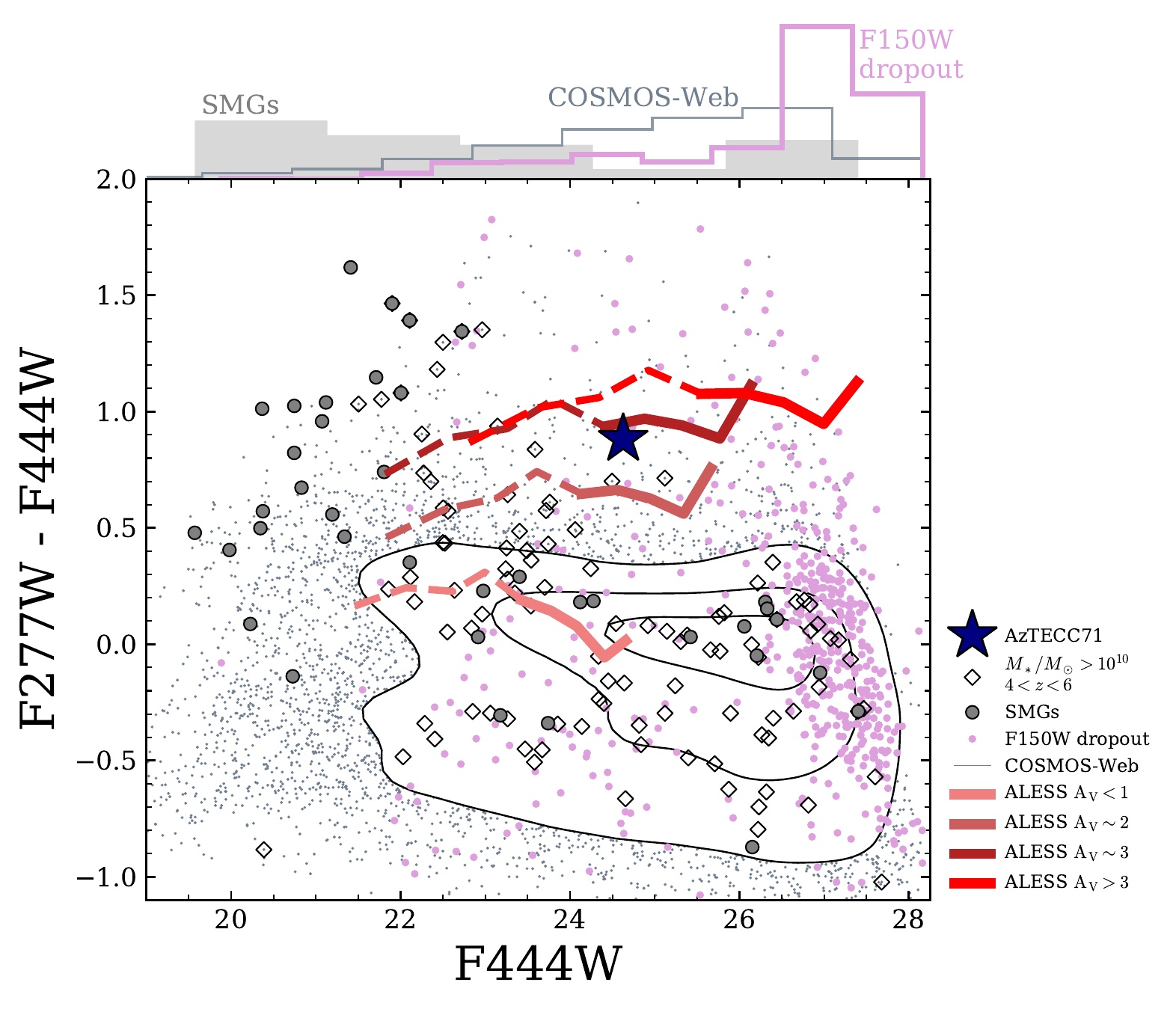}
    \caption{F444W/F277W color-magnitude diagram for galaxies in COSMOS-Web. We show every galaxy over the 77 arcmin$^2$ area in the background, with contours drawn at 16th, 50th, and 84th percentiles. From within that sample, we highlight: F444W sources that drop out of the F150W filter (purple circles), and $4<z<6,\,\,\mathrm{M_*/M_\odot>10^{10}}$ galaxies with counterparts in COSMOS2020 (diamonds, \citealt{Weaver2022}). SMGs with 3 GHz counterparts are shown with dark grey circles. Red tracks correspond to redshifted empirical SEDs between $z=2-4$ (dashed) and $z=4-6$ (solid) from the ALESS sample \citep{daCunha2015} at progressively higher $A_V$. We normalize the ALESS tracks such that the F444W flux is approximately that of the averaged optically-faint ALESS SEDs for their mean redshift: $\sim2$ nJy for $\langle z\rangle\mathrm{_{OIR-faint}^{ALESS}}=3.7$ \citep{daCunha2015}. AzTECC71 (black star) is amongst the reddest population of galaxies found in the first six pointings of COSMOS-Web, is redder than known $4<z_{phot}<6$ massive galaxies from COSMOS2020, and is brighter in F444W than most objects that drop out of F150W. AzTECC71 is also the faintest F444W source with F277W-F444W$>0.2$ and an FIR counterpart confirmed with an ALMA or VLA detection. 
    }
    \label{fig:colormag}
\end{figure*}

\subsection{Implications for high-z star-forming galaxy populations}
The incidence of objects like AzTECC71 in upcoming \textit{JWST} surveys will provide a key test on obscured star-formation and the production of dust in the early Universe. 
We show in Figure \ref{fig:colormag} the F444W/F277W color-magnitude space occupied by galaxies in the COSMOS-Web January 2023 coverage \citep{Casey2022}. Starting with SCUBA-2 detections and validating counterparts with VLA 3 GHz imaging and then \textit{Spitzer} MIPS and IRAC counterparts, the bulk of the SMGs within the COSMOS-Web footprint are bright in \textit{Spitzer} and have F444W$\sim20-22$ magnitudes. AzTECC71 is redder than SMGs fainter in F444W as well as known $4<z_{phot}<6$ massive galaxies from COSMOS2020 \citep{Weaver2022}. Average SEDs binned by $A_V$ from ALESS and then redshifted to $z\sim5-6$ are consistent with AzTECC71's color and derived properties \citep{daCunha2015}, and suggest that some fraction of sources with F277W-F444W$\,>0.5$ and F444W$\,>24$ could be $z>5$ dusty, star-forming galaxies. 

There are 627 objects in the COSMOS-Web area analyzed in this work with F277W$-$F444W$\,>0.5$. Of these, 80 (13\%) drop out of the F150W filter and have F444W$\,>26$ magnitude, 15 (2\%) are SMGs with a radio counterpart, and we infer 21 (3\%) are $4<z<6$ and $\mathrm{M_*>10^{10}\,M_\odot}$ with photo-$z$'s from COSMOS2020. 300 (48\%) of these galaxies have no counterpart from COSMOS2020 within $1^{\prime\prime}$. Their division between a broader OIR-faint galaxy population and an OIR-faint FIR-bright (e.g., $S_{\nu,850 \rm \mu m}\gtrsim1\,{\rm mJy}$) one could change the inferred dust-obscured star-formation density at $z\sim4-6$. \cite{Barrufet2022} argue that \textit{HST}-dark \textit{JWST}-detected galaxies heavily obscured in the optical and with $\mathrm{
\log M_*/M_\odot<10.5}$ might dominate over the more massive SMGs at $z>5$, which could push the dust-obscured star-formation rate density to larger values than previously measured from bright 2mm sources \citep{Zavala2021,Casey2021}. We find four more objects like AzTECC71 with ALMA-counterparts to SCUBA-2 detections that also drop out of the F150W filter and have $z_{phot}>4$ by virtue of bright far-IR emission combined with very red F277W$-$F444W  (Manning et al., in prep.). Based on these sources (including AzTECC71) and the current area of COSMOS-Web's footprint (77 arcmin$^2$) we estimate a number density of $4<z<6$ dusty, star-forming galaxies with \loglir$\,>11$ of $n\sim10^{-5}\,\,\mathrm{Mpc^{-3}}$. This is high relative to the median reported in the literature ($10^{-5.5}\,\,\mathrm{Mpc^{-3}}$, as compiled by \citealt{Long2022}), which may require revised estimates on the depletion/quenching timescale for dusty star-formation in the first $\sim$Gyr of galaxy formation. However, the disparity in reported volume densities is largely due to different survey areas and wavelengths. Large area \textit{JWST} surveys like COSMOS-Web combined with deep and contiguous far-IR/sub-mm data will be critical for reducing these systematic sources of uncertainty. 

\section{Summary and Conclusion\label{sec:conclusion}}
We report the \textit{JWST}/NIRCam detection from the COSMOS-Web survey of AzTECC71, a known sub-mm source with no previous detection below $850$\micron. We identify counterparts in NIRCam/F277W and F444W using ALMA Band 6 imaging to localize the sub-mm emission. AzTECC71 is not detected in the COSMOS-Web F115W and F150W images, and is not detected in other ground- or space-based imaging below 2.7\micron. 
Based on multi-wavelength SED modeling, AzTECC71 is a massive ($\log\mathrm{M_*/M_\odot=10.8}$), and IR-luminous ($\log\mathrm{L_{IR}/L_\odot=12.75}$) galaxy with a high ($>99\%$) probability to be at $z>4$. 

AzTECC71 is broadly similar with respect to stellar properties and far-IR flux densities to known optically-dark/faint dusty, star-forming galaxies \citep{Williams2019,Talia2021,Manning2022}. This object is the faintest confirmed F444W counterpart to sub-millimeter galaxies in COSMOS-Web with SNR$_{850\rm \mu m}>4$ and F277W-F444W$\,>0.25$, and a member to a larger sample of high-redshift dusty, star-forming galaxies with no prior optical/near-infrared counterpart (Manning et al., in prep.). If ALMA-confirmed sub-millimeter sources that drop out of F150W in COSMOS-Web collectively fall at $z>4$ as is the case for AzTECC71, then the number counts of luminous, infrared galaxies at $4<z<6$ would be 0.5 dex higher than the median reported in the literature \citep{Long2022}. This could require revised estimates on the $z>4$ IR star-formation rate density, as well as for the quenching timescale of dusty, star-forming galaxies at this epoch. 

The hunt for and characterization of optically-faint far-infrared bright galaxies stands on a precipice. The combination of \textit{JWST}, ALMA and the VLA has significantly improved our capacity to find counterparts to bright far-IR sources and characterize the stellar populations in high-redshift ($z>4$) dusty, star-forming galaxies. In parallel, the upcoming generation of deep imaging surveys with \textit{JWST} are finding very faint optical/near-IR sources that plausibly occupy the LIRG regime at $z>4-6$ but lack IR coverage necessary to robustly confirm or deny a high dust-obscured star-formation rate. At this epoch the incidence of dusty star-forming galaxies remains unconstrained owing to limiting sensitivities of ground-based far-IR facilities that struggle to survey below the ULIRG limit at $z>4$, though there is some evidence for more obscured star-formation than previously thought at this epoch \citep{Wang2016,Gruppioni2020,Talia2021,PerezGonzalez2022,Barrufet2022,Rodighiero2023}. 
Moreover, the ``optically-faint'' classification is becoming a confusing identifier as the population diversity of objects discovered for the first time in the optical by \textit{JWST} grows. 
Upcoming instruments such as ToLTEC on the LMT will push to deeper \lir\ limits with $\sim5^{\prime\prime}$ resolution at 1.1 mm, enabling better counterpart matching between large optical/near-IR and far-IR surveys. 
With these state-of-the-art data sets, OIR-faint/dark galaxies should distinguish between far-IR bright or faint populations in classification schemes where possible. 
Even then, a cold far-infrared space telescope with either a large aperture of sufficient spectroscopic sensitivity to disentangle confused sources along the frequency axis is needed fully uncover the very distant IR Universe.

\begin{acknowledgments}
JM would like to thank C.~Williams for making the SED of COS-3mm-1 readily available. This work is based [in part] on observations made with the NASA/ESA/CSA \textit{JWST}. The data were obtained from the Mikulski Archive for Space Telescopes at the Space Telescope Science Institute, which is operated by the Association of Universities for Research in Astronomy, Inc., under NASA contract NAS 5-03127 for \textit{JWST}. These observations are associated with program ID $1727$. Support for program 1727 was provided by NASA through a grant from the Space Telescope Science Institute, which is operated by the Association of Universities for Research in Astronomy, Inc., under NASA contract NAS 5-03127.
This paper makes use of the following ALMA data: ADS/JAO.ALMA:2013.1.00118.S. ALMA is a partnership of ESO (representing its member states), NSF (USA) and NINS (Japan), together with NRC (Canada), MOST and ASIAA (Taiwan), and KASI (Republic of Korea), in cooperation with the Republic of Chile. The Joint ALMA Observatory is operated by ESO, AUI/NRAO and NAOJ. The National Radio Astronomy Observatory is a facility of the National Science Foundation operated under cooperative agreement by Associated Universities, Inc.
This work used the CANDIDE computer system at the IAP supported by grants from the PNCG, CNES, and the DIM-ACAV and maintained by S. Rouberol.
\end{acknowledgments}

\bibliography{references}
\bibliographystyle{aasjournal}

\end{document}